\documentclass[aps,prl,twocolumn,superscriptaddress]{revtex4-2}

\usepackage{dsfont}
\usepackage{float}
\usepackage{xcolor}
\usepackage{physics}
\usepackage[caption=false]{subfig}

\usepackage{hyperref}
\hypersetup{
	breaklinks=true,
	colorlinks=true,
	allcolors=blue,
}

\usepackage{txfonts}
\usepackage{amsmath}
\usepackage{amssymb}
\usepackage{graphicx}
\usepackage{dcolumn}
\usepackage{bm}
\usepackage{array}
\usepackage{multirow}

\begin{document}
	
    \title{Phase diagram of lasing under correlated pump from GPU-accelerated Truncated Wigner dynamics}

 	\author{Oksana Chelpanova\href{https://orcid.org/0000-0002-1679-1359}{\includegraphics[height=1.7ex]{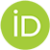}}}
        \email{oksana.chelpanova@weizmann.ac.il}
\affiliation{Department of Physics, The State University of New York at Buffalo, Buffalo, NY 14260, USA}

\author{Martino Stefanini\href{https://orcid.org/0000-0002-8095-9603}{\includegraphics[height=1.7ex]{orcid-logo.png}}}%
\affiliation{Department of Physics, The State University of New York at Buffalo, Buffalo, NY 14260, USA}
\affiliation{%
ICTP South American Institute for Fundamental Research and
Instituto de Física Teórica, UNESP - Univ. Estadual Paulista,
Rua Dr. Bento Teobaldo Ferraz 271, 01140-070, São Paulo, SP, Brazil
}%

\author{Michael F. O’Keeﬀe\href{https://orcid.org/0009-0006-7803-8051}{\includegraphics[height=1.7ex]{orcid-logo.png}}}
\affiliation{NVIDIA Corporation, 2788 San Tomas Expressway, Santa Clara, 95051, CA, USA}

	\author{Jamir Marino\href{https://orcid.org/0000-0003-2585-2886}{\includegraphics[height=1.7ex]{orcid-logo.png}}}
\affiliation{Department of Physics, The State University of New York at Buffalo, Buffalo, NY 14260, USA}

	\begin{abstract}
Superradiant (SR) lasers store optical coherence in the atomic medium rather than the cavity field, but the incoherent drive that sustains inversion imposes a trade-off: local pumping yields coherent light at a rate that grows linearly with the atom number $N$, heating the medium through photon recoil, whereas fully collective pumping removes this scaling but limits the emission to partial coherence. We interpolate between these limits employing a spatially correlated pump on a chain of $N$ two-level atoms, with rates decaying  with distance between atoms as a power law of exponent $\alpha$. 
To study systems beyond the reach of exact solutions, we employ the Truncated Wigner Approximation (TWA), whose independent trajectories are ideally suited to GPU parallelism.
Harnessing this, we perform a full scan of the steady-state observables for up to $10^4$ atoms at a computational cost that is practical.
Our findings indicate that ultra-narrow emission persists for all $\alpha$, while the coherence improves as the pump becomes shorter ranged, with $g^{(2)} \to 1$ surviving at least down to $\alpha \approx 1$, indicating that fully coherent light thus does not require local pumping.
The drive strength needed for lasing is reduced by a factor
$N^{1-\alpha}$ for $\alpha < 1$, and by $\log N$ as $\alpha \to 1$,
parametrically suppressing recoil heating; notably, $\alpha = 1$ matches the far-field envelope of dissipative couplings in free space. The correlation range
of the pump thus acts as a knob trading drive intensity, and the heating it
causes, against optical coherence.
\end{abstract}
	\maketitle
	 

     \begin{figure}[b]
    \centering
    \includegraphics[width=0.8\linewidth]{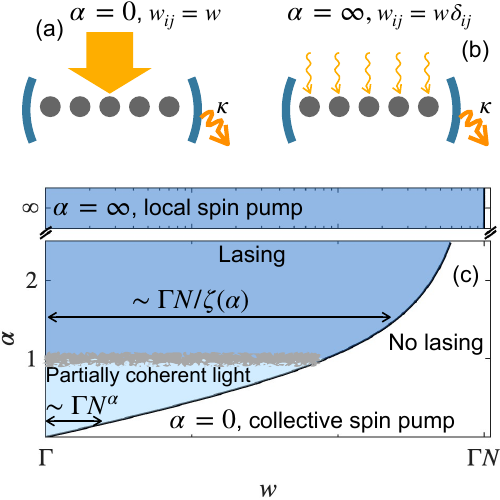}
    \caption{Schematics of the setup for collective (a) and local (b) spin pump. (c) Sketch of the phase diagram that identifies parameter regimes where atoms emit coherent/partially coherent light (lasing, blue) vs thermal radiation (white) as a function of pump rate and the exponent $\alpha$ which controls the range of correlated pump $w_{ij}=w/(1+|i-j|)^{\alpha}$. Here $w$ is plotted in logarithmic scale. We identify the upper bound on the transition from partially coherent to coherent light at $\alpha\lesssim 1$ which allows to reduce the pump intensity by at least factor $\log N$ compared to conventional SR lasing. 
    }
    \label{fig:schematics}
\end{figure}

\textit{\textbf{Introduction}---}Lasers with ultra-narrow linewidths and long-term phase stability are central to precision metrology, from optical atomic clocks~\cite{nicholson2015systematic,chen2009active} to gravitational-wave detection~\cite{bothwell2022resolving,Haroche_laser}. In conventional lasers, thermal fluctuations of the cavity mirrors ultimately limit   phase stability~\cite{PhysRevLett.101.260602,vibrations_1,cole2013tenfold,PhysRevA.99.061801,PhysRevX.13.041002}. On the other hand, superradiant (SR) lasers operating in the bad-cavity regime circumvent this limitation by storing coherence in the atomic medium rather than in the cavity field: the emission frequency is pinned to the atomic transition, and cavity-induced noise is strongly suppressed~\cite{PhysRevX.8.021036,Holland_SR_laser,PhysRevResearch.7.013292}. In its minimal form, an SR laser involves only two incoherent processes---collective decay through the lossy cavity mode and incoherent pumping that maintains population inversion~\cite{Holland_SR_laser}.

The main limitation comes from the pump. A local pump must operate at a rate that scales linearly with the atom number $N$ to reach the optimal lasing regime~\cite{Holland_SR_laser}, while the accompanying photon recoil heats the atomic medium and degrades the laser performance~\cite{Thompson_laser,reilly2025fullycollectivesuperradiantlasing}. Making the pump fully collective removes this $N$ scaling, but the resulting symmetry limits the coherence to $g^{(2)}\to 6/5$ at a fine-tuned operating point~\cite{sm,reilly2025fullycollectivesuperradiantlasing}. Restoring $g^{(2)}\to 1$ within a collective pumping scheme has so far required a third atomic level~\cite{reilly2025fullycollectivesuperradiantlasing}. A complementary approach is spatially selective pumping, in which only a subensemble of atoms is driven while the remaining atoms are passive. The interplay among coherent dipole--dipole exchange, collective loss, and inhomogeneous pumping then acts as an interference mechanism that filters the emission, yielding coherent, ultra-narrow light~\cite{chelpanova2026knobtuneallphasecontrolled,Ritsch_partial_laser,Bruder_laser,bychek2026continuousnarrowlinewidthsuperradiancewaveguide}. These schemes, however, rely on coherent interactions between the subensembles, while the drive applied to the pumped atoms remains strong, so recoil heating persists. Existing strategies therefore trade reduced recoil heating against optical coherence.

In this Letter, we consider incoherent  correlated pump  to connect the local
and collective limits directly. We consider a chain of $N$ two-level atoms subject to collective loss mediated
by the bad cavity, and to incoherent pumping with pump rates
$w_{ij}=w/(|i-j|+1)^{\alpha}$, where the exponent $\alpha$ interpolates
continuously between collective [$\alpha=0$,
cf.~Fig.~\ref{fig:schematics}(a)] and local [$\alpha\to\infty$,
cf.~Fig.~\ref{fig:schematics}(b)] pumping. Any intermediate
$0<\alpha<\infty$ breaks permutation invariance, so no exact solution is
available: correlated pumping bridges the two solvable limits through a
regime that has so far remained uncharted. 
Using GPU-accelerated dissipative
truncated Wigner approximation (TWA)~\cite{1wwv-k7hg}, we evaluated steady state observables---magnetization, intensity, second-order coherence, and
linewidth---for systems of up to $10^4$ atoms. Our aim is not to model a specific platform, but to isolate a question the
two solvable limits cannot answer: which features of an incoherent pump
control the three defining properties of a superradiant laser --- the drive
strength required to lase, the coherence of the emitted light, and the
linewidth? In our setting these features are encoded in the spectrum of the
pump-rate matrix $w_{ij}$, which interpolates from rank one (collective) to
full rank (local) as $\alpha$ grows; the power law is the minimal
one-parameter family that tunes this spectrum continuously, and should be
regarded as a theoretical laboratory rather than as the microscopic kernel
of a particular implementation. Studying the impact of our findings in more realistic setups such as including dipolar interactions, the full profile of $w_{ij}$ associated to a specific light-matter implementation, and so on, is the purpose of forthcoming work.

Our central result is that correlated pumping of power-law type directly relaxes the drive
requirement of the SR laser, cf. Fig.~\ref{fig:schematics}(c). Relative to the standard local-pump value, the
rate needed to reach optimal lasing is reduced by a factor $\sim
N^{1-\alpha}$ for $\alpha<1$, by $\log N$ at $\alpha=1$, and by
$\zeta(\alpha)$ for $\alpha>1$ [$\zeta(\alpha)$ being the Riemann zeta function], parametrically suppressing the recoil heating that limits current
implementations~\cite{Thompson_laser}. This reduction does not compromise the
defining feature of the SR laser: superradiant emission with an ultra-narrow
linewidth  persists for \emph{all} values of
$\alpha$. What it costs is coherence, which improves as the pump becomes
shorter ranged: fully coherent emission, $g^{(2)}\to1$, survives at least
down to $\alpha\approx1$, and our finite-size extrapolation suggests it may
extend below. In this sense $\alpha$ acts as a spectroscopic dial for the pump matrix,
revealing that the three properties decouple: the drive requirement tracks
the largest eigenvalue of $w_{ij}$, coherence tracks how strongly the pump
breaks permutation symmetry, and the ultra-narrow linewidth --- inherited
from the collective loss showing minor sensitivity to the pump range.

Beyond this application, GPU acceleration substantially lowers the cost of
constructing and analyzing a full steady-state phase diagram compared with the CPU implementations typically used for TWA, allowing us to characterize the steady state throughout the
parameter regime connecting the two permutation-symmetric solvable limits.
This establishes GPU-accelerated TWA as a practical tool for
driven-dissipative spin systems beyond permutation symmetry.

\textit{\textbf{Model}---}As a minimal model, we consider the standard setup of Ref.~\cite{Holland_SR_laser}: a one-dimensional (1d) chain of $N$ atomic dipoles, modeled as two-level atoms, coupled uniformly to a single cavity mode $a$ with interaction strength $g$. We represent the two-level atoms with (pseudo)-spins-$1/2$ using the Pauli matrices $\sigma_i^{\mu}$, with $i=1,\ldots,N$, $\mu=\{x,y,z\}$, and $[\sigma_i^\mu,\sigma_j^\nu]=2\mathrm{i}\epsilon_{\mu\nu\eta}\delta_{ij}\sigma_i^\eta$. We further introduce the raising and lowering operators $\sigma_i^{\pm}=(\sigma_i^x\pm i\sigma_i^y)/2$ and the collective spin operators $S^{\mu}=\sum_i \sigma_i^{\mu}$. The atoms are incoherently pumped at rate $w$ and undergo local decay at rate $\gamma$, which is much smaller than the other incoherent rates and is therefore neglected. The cavity mode $a$ decays at rate $\kappa$, which is the largest energy scale in the model. In this bad-cavity regime, the cavity field is enslaved to the spins and can be integrated out, resulting in collective spin dissipation at the effective rate $\Gamma=4 g^2/\kappa$. Focusing on a minimal model and assuming a small atom--cavity detuning, we also neglect the coherent interaction between atoms,   mediated by the same cavity field.

The effective spin dynamics in the co-rotating frame is then governed by the Lindblad master equation~\cite{breuer2002theory,stefanini2025lindblad,fazio2025many}
\begin{equation}\label{eq:model}
\begin{aligned}
    \dot\rho= \mathcal L\rho&=\frac {\Gamma}{2}
     \left(2 S^-\rho S^+- \{S^+ S^-,\rho\}  \right)\\
     &+\sum_{ij}\frac{w_{ij}}{2}\left(2\sigma _j^+\rho\sigma_i^- - \{\sigma_i^-\sigma_j^+,\rho\}\right).
\end{aligned}
\end{equation}
This model captures the competition between collective spin loss induced by coupling to the bad-cavity mode and incoherent pumping with rates $w_{ij}$, which drives population inversion. In the spirit of providing a minimal model that interpolates between lasing with local  and global pumps, we assume a power-law profile for the pump rates,
\begin{equation}\label{eq:corr_pump}
    w_{ij}=\dfrac{w}{\left(|i-j|+1\right)^{\alpha}}.
\end{equation}
The exponent $\alpha$ interpolates between collective pumping at $\alpha=0$ [$w_{ij}=w$, Fig.~\ref{fig:schematics}(a)] and the local pumping of the standard SR laser in the limit $\alpha\to\infty$ [$w_{ij}=w\delta_{ij}$, Fig.~\ref{fig:schematics}(b)]. Finite values of $\alpha$ describe the intermediate regime of spatially correlated pumping. Physical mechanisms that correlate incoherent channels --- repumping through
long-wavelength transitions, or emission into free space and waveguides
~\cite{AnaAG_earh_like,PhysRevLett.125.263601,gavalda2026fourierimagingcollectivespontaneous,novotny2012principles} --- generically produce kernels with additional structure:
oscillatory phases, polarization dependence, and an accompanying coherent
dipole--dipole exchange. Eq.~\eqref{eq:corr_pump} deliberately 
keeps only a positive, monotonically decaying envelope with a single
tunable range $\alpha$ -- the minimal setting in which the role of pump
correlations can be isolated. Which microscopic schemes realize a given
effective $\alpha$, and how the neglected structure modifies the picture,
are separate questions to which we return in forthcoming work.

The correlated pump in Eq.~\eqref{eq:corr_pump} can be decomposed into $N$ effective dissipation channels, with effective pump rates given by the eigenvalues of the pump-rate matrix, $\{w_{\nu}\}=\mathrm{eig}(w_{ij})$~\cite{PhysRevLett.125.263601}. The range of pump rates spanned by the lasing region is set by the largest eigenvalue $w_{\nu}^{\max}$, which for large $N$ is approximately given by the largest eigenvalue, namely proportional to the inverse Kac factor $K^{-1}(N,\alpha)=\sum_j w_{ij}$~\cite{CAMPA200957,defenu,Defenu_pnas,PhysRevB.106.224308}, hence
\begin{equation}\label{eq:max_eig}
    w_{\nu}^{\max}\sim
    \left\{  
    \begin{aligned}
        &w N^{1-\alpha},\quad &\alpha<1,\\
        &w\log(N), \quad &\alpha=1,\\
        &w\zeta(\alpha), \quad &\alpha>1,\\
    \end{aligned}
    \right.
\end{equation}
where $\zeta(\alpha)$ is the Riemann zeta function \footnote{Notice that the divergence of $\zeta(\alpha)$ at $\alpha\to1^+$ is cutoff by system size, since $w_\nu^\textup{max}$ converges to $ w \zeta(\alpha)$ only if $\alpha-1\gg(\log N)^{-1}$, otherwise it has the same scaling as $\alpha=1$, $w_\nu^\textup{max}\sim w \log N$.}. To place all $\alpha$ regimes on a common scale, we rescale the pump rate as 
$\tilde w=w_{\nu}^{\max}/(\Gamma N)$, removing the trivial extension of the superradiant emission region with $N$. In practice, we obtain $w_{\nu}^{\max}$ by numerically diagonalizing the matrix $w_{ij}$ and taking its largest eigenvalue; we have verified that this agrees with the scaling predicted by Eq.~\eqref{eq:max_eig} with numerical coefficients depending on boundary conditions considered. In this Letter, we use open boundary conditions. 

The smallest eigenvalue, in turn, sets a lower bound on the relaxation time needed to reach the steady state, $\tau_r\ge 1/w_{\nu}^{\min}$. For a power-law profile it can be approximated~\cite{gray2006toeplitz} as $w_{\nu}^{\min} \approx w(2 \eta(\alpha)-1)$, where $\eta(\alpha)=\sum_{m=1}^{\infty} (-1)^{m-1}/m^{\alpha}$ is the Dirichlet eta function. Close to the collective limit $\alpha\to 0$, the relaxation time scales as $\tau_r\sim (w\alpha)^{-1}$, while for $\alpha\to \infty$ it saturates to $w^{-1}$.

\textit{\textbf{Observables and method}---}To distinguish the SR lasing regime from other phases, we evaluate a standard set of observables in the steady state~\cite{walls2008quantum,reilly2025fullycollectivesuperradiantlasing}: the steady-state magnetization $\langle S^z\rangle$, the intensity $\langle S^+ S^-\rangle$, the second-order coherence at zero delay, $g^{(2)}\equiv g^{(2)}(0)=\langle S^+ S^+ S^- S^-\rangle/\langle S^+ S^-\rangle^2$, and the linewidth $\Delta\nu$, extracted as the full width at half maximum of the spectral function $S(\omega)=2\Re \int dt\, e^{\mathrm{i}\omega t}\,
\langle S^+(t)S^-(0)\rangle$. In the SR lasing regime, $\langle S^+S^-\rangle\propto N^2$, $g^{(2)}\to1$, and the linewidth becomes ultra-narrow, $\Delta\nu\sim\Gamma$, without any additional scaling with the system size $N$~\cite{Holland_SR_laser,Holland_g2}.

Exact solutions for large $N$ are available only for values of $\alpha$ that preserve permutation invariance~\cite{lloyd2026permutationsymmetricquantumtrajectories,PIQS,barberena2025generalizedholsteinprimakoffmapping1n,PhysRevA.87.062101,9p44-kwfl}.  For $\alpha\notin \{0,\infty\}$, the Hilbert space grows exponentially with system size, and we must rely on approximate methods. In this work, we use the TWA~\cite{1wwv-k7hg,GUARDIOLANAVARRETE202587,PhysRevResearch.4.043136,hartmann2026truncatedwignerapproximationspins,noel2026}, which provides a qualitatively reliable description in the parameter regimes considered below (and quantitatively correct results for $\tilde w\lesssim 1/4$), with known quantitative limitations at strong pumping, whereas second-order cumulant expansions~\cite{Plankensteiner2022quantumcumulantsjl} can yield unphysical values of $g^{(2)}$ at weak pumping~\cite{Holland_g2}---see benchmark in supplemental materials (SM)~\cite{sm}. 

\begin{figure}
    \centering
    \includegraphics[width=0.9\linewidth]{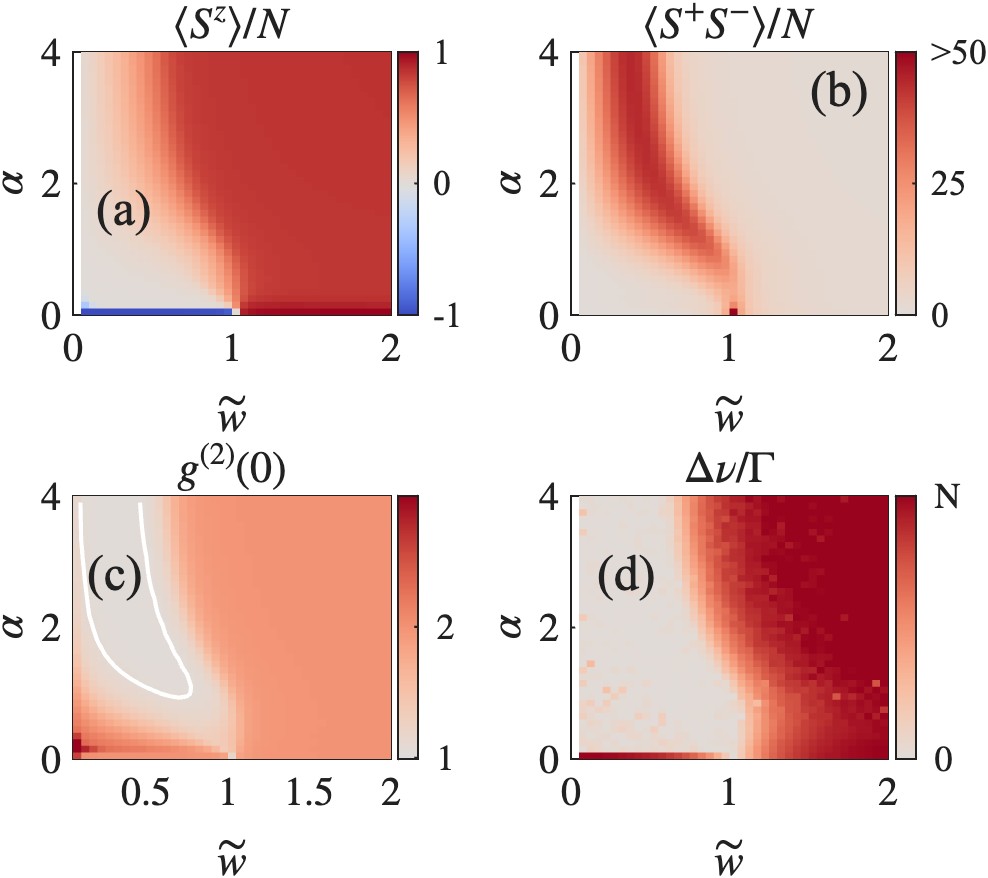}
    \caption{Steady-state observables: (a) magnetization $\langle S^z\rangle$, (b) intensity $\langle S^+ S^-\rangle$, (c) second-order coherence $g^{(2)}$, and (d) linewidth $\Delta\nu$ as functions of the normalized pump rate $\tilde w$ and the exponent $\alpha$. The white contour in panel (c) delimits an approximate region of coherent emission. 
    For $\alpha=0$, the pump and loss preserve strong permutation symmetry, which confines the dynamics to the symmetric Dicke manifold and precludes an extended coherent-lasing regime. For $\alpha \gtrsim 2$, the pump becomes effectively short-ranged.
   }\label{fig:500_spins}
\end{figure}

For each simulation discussed in this paper, we evolve the stochastic dynamics of a system of $N$ spins over $N_{\mathrm{trj}} = 2^{13} = \mathcal{O}(10^{4})$ trajectories using a semi-implicit integration scheme~\cite{Mentink_2010a}; the equations of motion are provided in the SM~\cite{sm}, and a detailed implementation of spin-preserving integration schemes can be found in Ref.~\cite{noel2026quantumclassicalrelaxationdynamics}. For each trajectory, we sample the initial conditions from a discrete distribution satisfying $\langle \sigma_i^{x,y}\rangle=0$ and $\langle \sigma_i^z\rangle =1$, and evolve the stochastic dynamics for a time $\tau_s$ until the system reaches the steady state. To keep the computational cost finite, we set $\tau_s=10/\Gamma$ for $\alpha<1$ and $\tau_s=10/w_{\nu}^{\min}$ for $\alpha\ge 1$, unless specified otherwise. We choose the time step $\Delta t$ such that $\max(\Gamma N, w_{\nu}^{\max})\cdot \Delta t<0.05$. We then evolve the system for an additional time $\tau=5/\Gamma$ and evaluate the observables by averaging the corresponding quantities over trajectories. The detailed procedure, including the equations of motion and the evaluation of $g^{(2)}$ and the linewidth $\Delta\nu$, is provided in the SM~\cite{sm}. We use NVIDIA A16, V100, A100, H100, and GH200 GPUs to evaluate the dynamics efficiently, taking advantage of the high degree of parallelism of the problem; further details are discussed in the End Matter.

\textit{\textbf{Results}---}In Fig.~\ref{fig:500_spins}, we plot the steady-state observables computed using the TWA as functions of $\tilde{w}$ and $\alpha$ for $N=500$, illustrating how the fine-tuned point of partially coherent emission at $\alpha=0$ evolves into the broad SR lasing region as $\alpha\to\infty$. The magnetization in panel (a) increases with the pump rate for all values of $\alpha$, although two distinct behaviors can be identified. For $\alpha>0$, $\langle S^z\rangle$ remains non-negative because local spin loss is neglected. By contrast, in the collective limit ($\alpha=0$), the competition between collective pump and collective loss produces a trigger-like transition at $\tilde w=1$, separating a negative-magnetization regime $\langle S^z\rangle\to -N$ for $\tilde w<1$ from the inverted state, $\langle S^z\rangle\to N$, for $\tilde w>1$ (see the SM~\cite{sm}). For $\alpha>0$, the magnetization crosses over from $\langle S^z\rangle\approx0$ to the inverted state as the pump rate approaches $\tilde w\approx1$. For $\alpha \gtrsim 1$ the TWA predicts this crossover at a slightly lower pump rate than the exact solution  (see the SM~\cite{sm}), because it overestimates the pumping noise in the short-ranged strong-pump regime. Overestimation of a noise strength in a spin-polarized state is a known feature of the TWA for local spin loss or pump in interacting systems~\cite{1wwv-k7hg,Mink_2023}, which can be cured in a single-body case~\cite{noel2026}. Nevertheless, the results remain qualitatively correct. The collapse of the crossover onto $\tilde w\approx1$ for all $\alpha$ confirms that the largest eigenmode of the pump matrix, Eq.~\eqref{eq:max_eig}, determines the extent of the lasing region.

The intensity in panel (b) shows that, for all values of $\alpha$, there is a range of pump rates for which $\langle S^+ S^-\rangle >N$, signaling a superradiant regime, since $\langle S^+S^-\rangle\propto N$ is expected for independent emitters~\cite{breuer2002theory}. By varying $N$, we confirm the $N^2$ scaling characteristic of Dicke superradiance~\cite{PhysRevResearch.4.023207}; see also the SM~\cite{sm}. The extent of this SR region grows with $\alpha$ for $\alpha<1$ and saturates in the short-range regime for $\alpha\gtrsim 2$. The second-order coherence in panel (c) reveals regions of thermal $g^{(2)}\to 2$, partially coherent  $1<g^{(2)}< 2$, and coherent emission  $g^{(2)}\to 1$, cf. Ref.~\cite{walls2008quantum}. The white contour delimits the region with $g^{(2)}\le 1.08$ and serves as a guide to the eye for the coherent-light regime at fixed $N$, with the cutoff value $1.08$ chosen to account for finite-size effects. 
Panel (d) shows that the linewidth is suppressed for $\tilde w\lesssim 1$ and broadens substantially at larger pump rates.
We observe an ultra-narrow linewidth regime for every value of $\alpha$ studied. The low-frequency resolution is limited by the time interval over which the dynamics is evolved after reaching the steady state and can be improved by increasing the total evolution time.

A striking feature of Fig.~\ref{fig:500_spins} is the qualitative difference between the fully collective limit $\alpha=0$ and any finite $\alpha$. At $\alpha=0$, both the pump and the loss preserve strong permutation symmetry, so the dynamics remain confined to the symmetric Dicke manifold with fixed total spin $J=N/2$. In this limit, the linewidth narrows only at the fine-tuned point $\tilde w=w/\Gamma=1$, where the Liouvillian acquires the additional symmetry $S^+\leftrightarrow S^-$~\cite{sm}. For any finite $\alpha$, the pump breaks strong permutation invariance and couples Dicke states belonging to sectors with different total spin $J$, allowing the dynamics to explore a larger portion of the Hilbert space. This enlargement of the accessible state space does not, by itself, imply existence of collective weak  $\mathrm{U}(1)$ symmetry breaking. 
for small $\alpha$, pumping occurs mainly through the collective channel [corresponding to the maximal eigenvalue~\eqref{eq:max_eig}], which yields only partial coherence (as in the $\alpha=0$ case).
Coherent emission emerges only at larger $\alpha$, where it signals synchronization of the atomic dipoles and is consistent with breaking of the collective weak $\mathrm{U}(1)$ symmetry in the thermodynamic limit.

To constrain the crossover from partially coherent to coherent emission, we analyze the second-order coherence $g^{(2)}$ across the parameter space. For fixed $N$ and $\alpha$, we vary $w$ and record the minimum value of $g^{(2)}$, denoted by $g^{(2)}_{\min}$. We sample $10$ equally spaced values of $\tilde w$ between $1/10$ and $1$. We restrict the analysis to $\alpha\gtrsim 0.5$, since the steady-state preparation time diverges as $(w\alpha)^{-1}$ in the collective limit.

\begin{figure}
    \centering
    \includegraphics[width=0.9\linewidth]{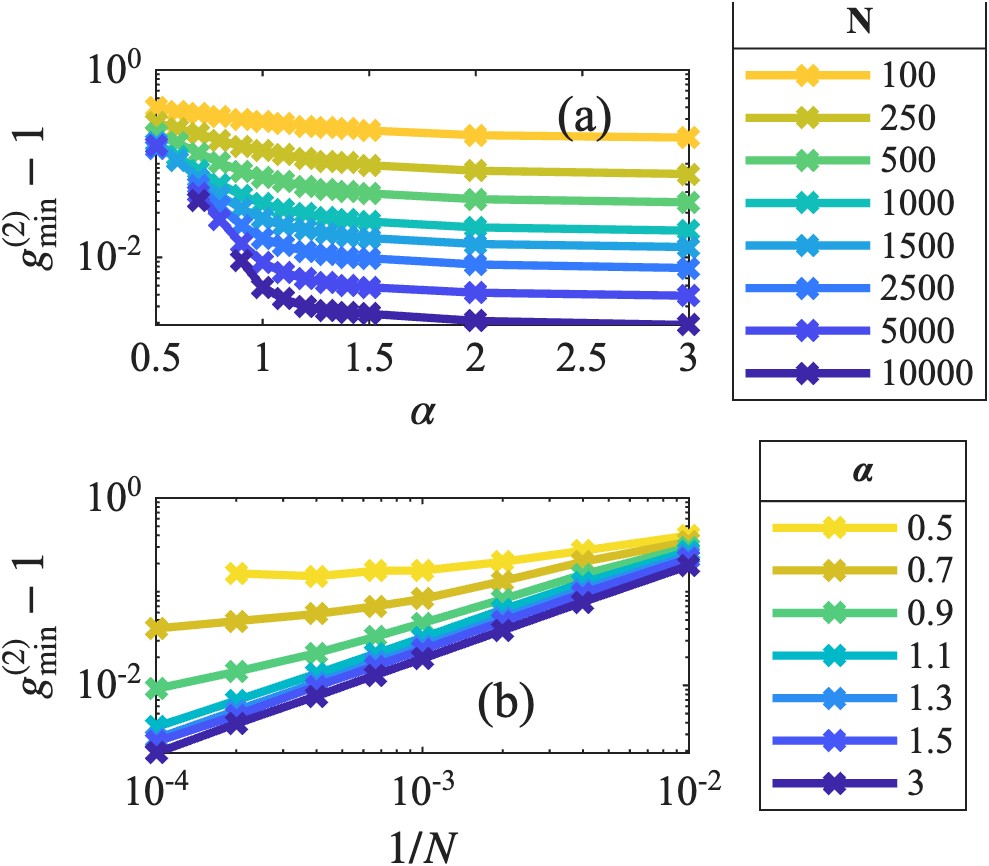}
  \caption{Minimal value of the second-order coherence as a function of $\alpha$ (a) and $N$ (b). The data place an upper bound near $\alpha\approx 1$ on the crossover to coherent emission and identify the region $\alpha>2$ as effectively local, or short-ranged, pumping. The $\alpha=0$ result, for which the second-order coherence approaches $g^{(2)}(0)\to 6/5$, is not shown here.}
    \label{fig:g2_scale}
\end{figure}

Figure~\ref{fig:g2_scale} shows $g^{(2)}_{\min}$ as a function of $\alpha$ in panel (a) and of $N$ in panel (b), with the system size varied from $10^2$ to $10^4$ spins. 
Even in the long-range regime $\alpha<1$, the coherence improves substantially as $\alpha$ and $N$ increase. Near the collective limit, at $\alpha\approx 0.5$, the improvement with system size is weak. At larger $\alpha$, however, $g^{(2)}_{\min}$ decreases more clearly with increasing $N$. A $1/N$ fit to the data in panel (b) suggests that, as $N\to\infty$, $g^{(2)}_{\min}\to 1.15$ for $\alpha=0.5$ and $g^{(2)}_{\min}\to 1.04$ for $\alpha=0.7$. For $\alpha\ge 0.9$, the extrapolated value lies even closer to the coherent limit $g^{(2)}=1$.

The optimal lasing region, characterized by the narrowest linewidth, occurs when the effective pump is weaker than the effective spin loss, which is set by $\Gamma N$. The extent of this region in terms of the unnormalized pump rate $w$ is therefore of order $\Gamma N/w_{\nu}^{\max}$, namely, $N^\alpha$ for $\alpha<1$ and $N/\zeta(\alpha)$ for $\alpha>1$. The latter scaling explains the broad lasing region recovered in the local-pump limit, where $\zeta(\alpha)\to 1$ as $\alpha\to\infty$. This result has a direct practical implication for the auxiliary lasers that mediate incoherent repumping. In the standard local-pump case, their intensity must scale with $N$, whereas correlated pumping relaxes this requirement: for $\alpha<1$, the required pump rate can be reduced by a factor $w^{\text{SR}}/w^{\text{cor}}\propto N^{1-\alpha}$, while for $\alpha>1$ it can still be reduced by a constant factor, thereby lowering both the noise and the recoil imparted to the atoms. This reduction comes at the cost of coherence: longer-range pumping requires a weaker drive but also increases $g^{(2)}$. The choice of $\alpha$ therefore controls the trade-off between the required drive intensity and the coherence of the emitted light, as summarized schematically in Fig.~\ref{fig:schematics}(c) [note that we use unnormalized pump rates in Fig.~\ref{fig:schematics}(c) and plot horizontal axis in logarithmic scale]. Interestingly, free-space emission, which can be used to implement the correlated pump, corresponds to an exponent $\alpha\approx 1$~\cite{novotny2012principles,PhysRevLett.125.263601} and already allows the pump rate to be reduced by a factor of order $\log N$ relative to the standard SR laser without the need for further dissipation engineering.

\textit{\textbf{Perspectives}}---Correlated dissipation opens a window onto new physics, with the potential to induce novel dynamical features and universality classes~\cite{PhysRevB.106.224308,Kushal1,Kushal2,Jamir_universality,prazeres2026kineticallyconstrainedsuperradiance}. It arises naturally in free-space and waveguide emission, and recent efforts have been directed toward engineering the geometry of atomic placement in order to explore a variety of correlation profiles~\cite{PhysRevLett.125.263601,
PhysRevResearch.4.023207,cardenaslopez2026steadystatespinordersuperradiance,PhysRevLett.134.173601,PhysRevB.111.064424,gunning2026interactiongeometrygroundstateproperties,periwal2021programmable}. A recent experiment~\cite{gavalda2026fourierimagingcollectivespontaneous} has demonstrated that different eigenmodes of the correlated decay matrix can be imaged and isolated in cold atomic clouds. The TWA appears to be a powerful tool for predicting the properties of such systems at low computational cost, while GPU acceleration enables significant speed-ups in simulating the dynamics of thousands of atoms. At the same time, the method must be applied with care, since a semiclassical approach cannot correctly describe intrinsically quantum systems that lack a well-defined mean-field order parameter~\cite{lxnv-d7ky}. In such cases, more sophisticated approaches are needed, such as a field-theoretical description that captures quantum effects more accurately~\cite{PhysRevResearch.6.043314}. 
Forthcoming work will promote the idealized
kernel of Eq.~\eqref{eq:corr_pump} to the dissipative profiles of concrete light--matter
platforms --- retaining the oscillatory structure and polarization
dependence of Green-tensor kernels, the accompanying coherent
dipole--dipole interactions, and geometries beyond the
one-dimensional chain --- so as to establish which implementations inherit
the drive reduction and the coherence identified here. Since the TWA
evolves an arbitrary pump matrix $w_{ij}$, or even correlated decay matrix of a Lindbladian, at unchanged computational cost,
the toolbox of this Letter including GPU acceleration carries over  these extensions.

\textit{\textbf{Data compliance}---}This data was produced by State University of New York
(SUNY) at Buffalo under Army Research Office (ARO) Award Number W911NF-26-1-A176. ARO,
as the Federal awarding agency, reserves a royalty-free, nonexclusive and irrevocable right to
reproduce, publish, or otherwise use this data for Federal purposes, and to authorize others to do
so in accordance with 2 CFR 200.315(b).

\textit{\textbf{Data availability}---}The data and code required to reproduce the results presented in this work are available
at~\cite{Chelpanova2026CorrelatedPumpData}.

\textit{\textbf{Acknowledgments}---}We thank N. Pancotti for contributing in the early stages of the project and Dori Sajdak for computational advisement and support. We acknowledge E. J. Davis and C. Rusconi for fruitful discussion. 
Research was sponsored by the Army
Research Office and was accomplished under Grant Number W911NF-26-1-A176. The views and
conclusions contained in this document are those of the authors and should not be interpreted as
representing the official policies, either expressed or implied, of the Army Research Office or the U.
S. Government. The U.S. Government is authorized to reproduce and distribute reprints for
Government purposes notwithstanding any copyright notation herein.
We gratefully acknowledge use of the research computing resources of the Empire AI Consortium, Inc, with support from Empire State Development of the State of New York, the Simons Foundation, and the Secunda Family Foundation. This work was performed in part at the University at Buffalo’s Center for
Computational Research~\cite{UBCCR}. This study was financed, in part, by the São Paulo Research Foundation (FAPESP), Brasil, Process Number 2024/22542-4. 
The authors acknowledge the use of generative AI tools in the course of this work. Code originally developed in MATLAB was translated to Python and partially vectorized and performance-optimized with AI assistance (Claude Opus 4.8, Anthropic; GPT-5, OpenAI and its previous versions); all resulting implementations were validated against the original code, and the results reported here were verified to be independent of the implementation. AI assistance was also used for data read-out and plotting scripts (panel layout, colormaps, scaling); no figure content was generated by AI. During manuscript preparation (autumn 2025–summer 2026), the authors used AI tools (Claude Opus 4.8 and Claude Opus 5 and Claude Fable 5, Anthropic; GPT-5.6 and earlier versions, OpenAI) for language editing, rephrasing, length control, typographical and consistency checking, and feedback on presentation and framing. AI tools were also used to identify relevant prior results in the literature; all such results are cited to their original sources. The physical analysis and interpretation are the authors' own. The authors take full responsibility for the content of the final manuscript.

\section*{End Matter}

\textit{\textbf{GPU acceleration}---}The TWA is highly parallelizable because the dynamics of individual trajectories evolve independently. For small system sizes and a moderate number of trajectories, the simulations can be performed efficiently on a personal computer. As the system size and the number of trajectories increase, however, it becomes advantageous to offload the computation to multicore CPUs or, more effectively, to GPUs. The main advantage of a GPU is its ability to execute a large number of computational threads concurrently, allowing many trajectories to be evolved in parallel. By contrast, even on a computing cluster, the number of CPU cores available to a single job is typically of order $10$--$100$. GPU acceleration can therefore substantially reduce the computational time required to obtain the semiclassical solution~\cite{tosca2025efficientvariationaldynamicsopen,SPIECHOWICZ2015140}.

In its simplest form, GPU acceleration is straightforward to implement. In Python, NumPy arrays can be replaced with CuPy~\cite{nishino2017cupy} or JAX~\cite{jax2018github} arrays, which are allocated on the device and support GPU-specific operations. In MATLAB, it is sufficient to replace arrays of type \texttt{double} with \texttt{gpuArray} objects~\cite{pct2024}. In this computational model, the CPU and GPU are referred to as the host and device, respectively. Arithmetic operations are performed on the device, whereas saving the results requires transferring the data back to the host. Because this transfer can introduce substantial overhead, it should be performed as infrequently as possible, ideally only once at the end of the simulation~\footnote{GPU usage can be profiled with NVIDIA Nsight Systems to analyze kernel scheduling and memory traffic. NVTX annotations in Python can be used to identify regions of the profiler output corresponding to specific operations in the code.}.

\begin{figure}
    \centering
    \includegraphics[width=\linewidth]{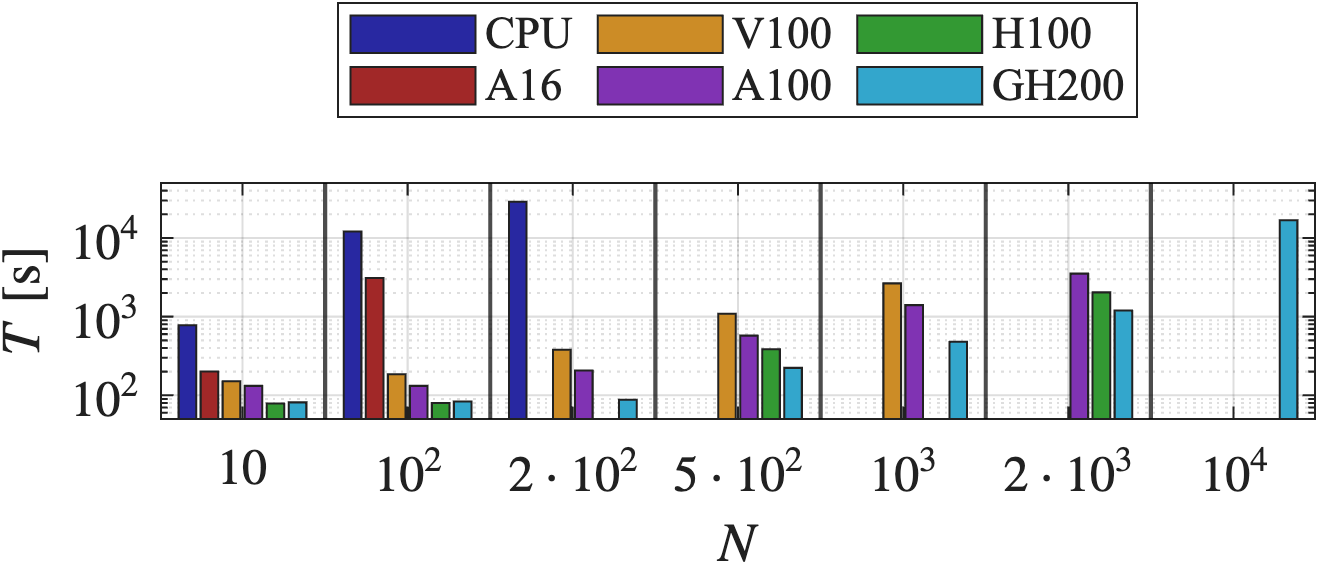}
    \caption{Computational time required to simulate the TWA dynamics of a system with correlated pumping as a function of the system size, using a single-core CPU and five NVIDIA GPUs. For all simulations, we fix $N_{\mathrm{trj}}=2^{13}$ and use the same number of integration steps, corresponding to a steady-state preparation time $\tau_r=5$ followed by an evolution time $\tau=10$. The decay rate $\Gamma$ is adjusted with $N$ so that $\Gamma N\Delta t$ remains constant. This protocol isolates the dependence of the computational cost on the system size. Simulations were terminated when the computation time exceeded several hours.}
    \label{fig:gpu}
\end{figure}

The computation times obtained using a single-core CPU and several NVIDIA GPUs---A16, V100, A100, H100, and GH200---are summarized in Fig.~\ref{fig:gpu}. Because the steady-state preparation time in our problem generally depends on $N$, we remove this dependence from the benchmark by fixing $\tau_r=5$. For each system size, we adjust the decay rate $\Gamma$ so that the largest decay rate $\Gamma N$ remains fixed. This allows all simulations to be performed with the same time step, satisfying $\Gamma N\cdot \Delta t=\mathrm{const}$. The benchmark therefore compares the time required to perform the same number of integration steps using the same simulation protocol at different system sizes.

Our numerical results show that, for $N_{\mathrm{trj}}=2^{13}$, the GPU simulations are faster than the single-core CPU simulation even for the smallest system considered, $N=10$. The CPU simulation requires $13$ minutes, whereas the same calculation takes $200$ seconds on the A16 and $78$ seconds on the GH200. As $N$ increases, the computational time on the CPU grows rapidly. Without multicore CPU parallelization, we obtained CPU results for $N\le 200$. Larger systems remain accessible in principle but require substantially longer computation times, and these simulations were therefore terminated. By contrast, the GPUs provide access to systems with $N\le 10^3$ within approximately one hour. More powerful devices, including the A100, H100, and GH200, allow still larger systems to be simulated. Using the GH200, we obtained results for $N=10^4$ spins, while even larger systems remain accessible at the cost of longer computation times. {The computational speedup depends on how the code is adapted for GPU execution. Nevertheless, GPU simulations are typically one to two orders of magnitude faster than the corresponding single-core CPU simulations, with the advantage increasing with system size until the available GPU memory becomes the limiting factor. Since trajectories are independent, a multi-threaded CPU implementation would narrow this gap roughly in proportion to the number of cores — typically 10–100 for a single cluster job — while a single GPU executes thousands of trajectories concurrently.}

Each data point in Fig.~\ref{fig:500_spins} was obtained in approximately $10$--$15$ minutes on the GH200. For simulations that do not require the evaluation of the linewidth, the total computation time can be reduced to a few seconds for $\alpha\gtrsim 2$. For smaller $\alpha$, however, the steady-state preparation time increases substantially.

\bigskip 

 \bibliography{lit}

\clearpage
\onecolumngrid

\setcounter{secnumdepth}{2}
\renewcommand{\thesection}{S\arabic{section}}
\renewcommand{\theequation}{S\arabic{equation}}
\renewcommand{\thefigure}{S\arabic{figure}}
\renewcommand{\thetable}{S\arabic{table}}
\setcounter{section}{0}
\setcounter{equation}{0}
\setcounter{figure}{0}
\setcounter{table}{0}

\title{Supplemental Material: Superradiant laser under correlated pump}

\maketitle

\section{Truncated Wigner approximation: dynamics and observables}
\label{sec:eom}

Below, we summarize the TWA procedure used to solve the dynamics with correlated pumping. Following Ref.~\cite{1wwv-k7hg}, the procedure consists of the following steps:
\begin{enumerate}
    \item For each quantum spin-$1/2$, $\mathbf{\sigma}_k=({\sigma}_k^x,{\sigma}_k^y,{\sigma}_k^z)$, we introduce a corresponding classical spin $\mathbf{s}_k=(s_k^x,s_k^y,s_k^z)$.

    \item The dynamics of each classical spin $\mathbf{s}_k$ is described by the following stochastic differential equations:
    \begin{equation}\label{eq:eom}
        \begin{aligned}
\partial_{t}s^{x}_{i}&=+\frac{1}{2}\Gamma s^{z}_{i}\sum_{j}s^{x}_{j}-\frac{1}{2}s^{z}_{i}\sum_{j}w_{ij}s^{x}_{j}+\xi^{x}_{i}s^{z}_{i}-\chi^{x}_{i}s^{z}_{i}\\
\partial_{t}s^{y}_{i}&=+\frac{1}{2}\Gamma s^{z}_{i}\sum_{j}s^{y}_{j}-\frac{1}{2}s^{z}_{i}\sum_{j}w_{ij}s^{y}_{j}+\xi^{y}_{i}s^{z}_{i}-\chi^{y}_{i}s^{z}_{i}\\
\partial_{t}s^{z}_{i}&=-\frac{1}{2}\Gamma\sum_{j}\left(s^{x}_{i}s^{x}_{j}+s^{y}_{i}s^{y}_{j}\right)+\frac{1}{2}\sum_{j}w_{ij}\left(s^{x}_{i}s^{x}_{j}+s^{y}_{i}s^{y}_{j}\right)-\left(\xi^{x}_{i}s^{x}_{i}+\xi^{y}_{i}s^{y}_{i}\right)+\left(\chi^{x}_{i}s^{x}_{i}+\chi^{y}_{i}s^{y}_{i}\right).
        \end{aligned}
    \end{equation}
The white noises satisfy $\langle \xi_i^{\alpha}(t)\xi_j^\beta(t') \rangle =\Gamma\delta(t-t')\delta_{\alpha\beta}$ and $\langle \chi_i^{\alpha}(t)\chi_j^\beta (t')\rangle =w_{ij}\delta(t-t')\delta_{\alpha\beta}$, where $\alpha,\beta=\{x,y\}$. These equations of motion can be interpreted as the evolution of a classical spin in an effective magnetic field, $\partial_{t}\mathbf{s}_{i}=\mathbf{s}_{i}\times \mathbf{b}^{eff}_{i}$, where the effective field $\mathbf{b}^{eff}_{i}$ contains both deterministic and stochastic contributions. The equations preserve the spin magnitude and can be solved efficiently using a semi-implicit integration scheme; see Ref.~\cite{Mentink_2010a}.

    \item We sample $N_{\mathrm{trj}}$ initial conditions for the classical spins from the discrete distribution $s_k^{x,y}=\pm 1$ and $s_k^z=1$, corresponding to a spin-polarized state with all emitters in the excited state.

    \item For the correlated-pump problem, we prepare the steady state by evolving the dynamics governed by Eq.~\eqref{eq:eom} for a time $\tau_s\approx 10/[w(2\eta(\alpha)-1)]$. This preparation time is used in the finite-size scaling analysis to account for the relaxation of all pump modes, including the slowest one. For the phase diagram shown in Fig.~2 of the main text, we instead use a shorter evolution time proportional to $1/\Gamma$ for $\alpha<1$. This is sufficient to resolve the qualitative regimes governed by the dominant pump mode, although the slowest modes may not be fully relaxed.

    \item The steady-state observables are evaluated by averaging the corresponding classical quantities over the unperturbed trajectories,
    $\langle \sigma_k^{\alpha}\rangle_{\mathrm{TWA}}=\frac{1}{N_{\mathrm{trj}}}\sum_i s_{k,i}^{\alpha}$,
    where $i=1,\ldots,N_{\mathrm{trj}}$. Symmetrically ordered two-point functions are similarly given by
    $\langle \sigma_k^{\alpha}\sigma_l^{\beta}\rangle _{\mathrm{TWA}}=\frac{1}{N_{\mathrm{trj}}}\sum_i s_{k,i}^{\alpha}s_{l,i}^{\beta}$.

    \item Single-time, normally ordered correlation functions can be expressed in terms of symmetrically ordered correlation functions by accounting for all possible permutations and using the commutation relations. In particular,
$\langle S^+ S^-\rangle = \langle S^+ S^-\rangle_{\mathrm{TWA} }+\langle S^z\rangle_{\mathrm{TWA}}/2$
and
\begin{equation}
\begin{aligned}
   \langle S^{+}S^{+}S^{-}S^{-}\rangle&=\langle  S^{+}S^{+}S^{-}S^{-} \rangle_{\mathrm{TWA}}+2\langle  S^{+}S^{z}S^{-}\rangle_{\mathrm{TWA}} \\
   &-\frac{1}{3}\langle S^{z} \rangle_{\mathrm{TWA}}+\frac{1}{2}\langle S^{z}S^{z}\rangle_{\mathrm{TWA}}-\frac{4}{3}\langle  S^{+}S^{-}\ \rangle_{\mathrm{TWA}}.
\end{aligned}
\end{equation}
For correlated emission, the following expression for the four-point function may also be required:
\begin{equation*}
    \begin{aligned}
\sum_{a,b,c,d}\Gamma_{ad}\Gamma_{bc}\langle\sigma_{a}^{+}\sigma_{b}^{+}\sigma_{c}^{-}\sigma_{d}^{-}\rangle 	&=\sum_{a,b,c,d}\left\langle \left(\sigma_{a}^{+}\Gamma_{ad}\sigma_{d}^{-}\right)\left(\sigma_{b}^{+}\Gamma_{bc}\sigma_{c}^{-}\right)\right\rangle_{\mathrm{TWA}} \\
	&+\sum_{a,b,c}\left(\Gamma_{aa}\left\langle \sigma_{b}^{+}\Gamma_{bc}\sigma_{c}^{-}\sigma_{a}^{z}\right\rangle_{\mathrm{TWA}} +\Gamma_{ab}\Gamma_{bc}\left\langle \sigma_{a}^{+}\sigma_{c}^{-}\sigma_{b}^{z}\right\rangle_{\mathrm{TWA}}\right)\\
	&-\frac{1}{3}\sum_{a}\Gamma_{aa}^{2}\langle \sigma_{a}^{z}\rangle_{\mathrm{TWA}}\\
	&+\frac{1}{4}\sum_{a,b}\left[\Gamma_{aa}\Gamma_{bb}\langle \sigma_{b}^{z}\sigma_{a}^{z}\rangle_{\mathrm{TWA}}+\Gamma_{ab}\Gamma_{ba}\langle \sigma_{b}^{z}\sigma_{a}^{z}\rangle_{\mathrm{TWA}}\right]\\
	&-\frac{4}{3}\sum_{a,b}\left(\Gamma_{aa}\Gamma_{ab}\left\langle \sigma_{a}^{+}\sigma_{b}^{-}\right\rangle_{\mathrm{TWA}} \right).
\end{aligned}
\end{equation*}
It is important to note that the TWA is a semiclassical approximation and does not capture all genuinely quantum contributions to higher-order correlation functions. Nevertheless, the method provides reliable estimates of the observables considered in the correlated-pump problem.
\end{enumerate}

To access two-time correlation functions, we follow Ref.~\cite{1wwv-k7hg} and decompose the correlation function into symmetric and antisymmetric parts:
\begin{equation}
\begin{aligned}
     \langle S^+(t) S^-(0)\rangle&=\frac{1}{4}\langle \{S^x(t)+\mathrm{i}S^y(t),S^x(0)-\mathrm{i}S^y(0)  \}\rangle\\
     &+\frac{1}{4}\langle [S^x(t)+\mathrm{i}S^y(t),S^x(0)-\mathrm{i}S^y(0)  ]\rangle.
\end{aligned}
\end{equation}
The first term is obtained directly within the TWA by averaging the product of classical variables, $\langle S^+(t)S^-(0)\rangle_{\mathrm{TWA}}$, over trajectories. The second term is evaluated using the Kubo formula~\cite{altland2010condensed}. We relate the commutator to the response function through $\theta(t)\langle [A(t), B(0)]\rangle =\mathrm{i}\chi_{AB}(t,0)$, so that the correction to the two-time correlation function is
$\mathrm{i}(\chi_{S^x S^x}(t,0)+\chi_{S^y S^y}(t,0)+\mathrm{i}\chi_{S^y S^x}(t,0)-\mathrm{i}\chi_{S^x S^y}(t,0))/4$.

To compute the response function, we apply an infinitesimal perturbation at time $t=0$, described by the Hamiltonian term $\delta H=J_B(t) B$ with $J_B=\epsilon_B\delta(t)$. Within linear response,
$\delta A(t)=\int \chi_{AB}(t,\tau)J_B(\tau)d\tau=\epsilon_B \chi_{AB}(t,0)$;
that is, one perturbs $B$ at $t=0$ and measures the resulting response of $A$.

More precisely, to evaluate the two-time correlation functions, we create two additional copies of all trajectories. In the first copy, we add an infinitesimal perturbation $\epsilon_x$ to the $x$ component of each spin,
$\mathbf{s}_k^{'}=(s_k^x+\epsilon_x,s_k^y,s_k^z)$.
In the second copy, we add an infinitesimal perturbation along the $y$ axis,
$\mathbf{s}_k^{''}=(s_k^x,s_k^y+\epsilon_y,s_k^z)$,
while leaving the remaining spin components unchanged. We then evolve the three copies of each trajectory for a time $\tau$, using identical realizations of the white noise in all three copies.

The two-time correlation functions are evaluated as
$\langle S^{\alpha}(t)S^{\beta}(0)\rangle =\frac{1}{N_{\mathrm{trj}}}\sum_{k,l}\sum_i s_{k,i}^{\alpha}(t)s_{l,i}^{\beta}(0) +\frac{i}{4N_{\mathrm{trj}}} \sum_i \chi_{S^{\alpha}S^{\beta}}^i$,
where the response functions are
$\chi_{S^{x}S^{x}}^i=\sum_{k} (s_{k,i}^{x'}(t)-s_{k,i}^x(t))/\epsilon_x$,
$\chi_{S^{x}S^{y}}^i=\sum_{k} (s_{k,i}^{x''}(t)-s_{k,i}^x(t))/\epsilon_y$,
$\chi_{S^{y}S^{x}}^i=\sum_{k} (s_{k,i}^{y'}(t)-s_{k,i}^y(t))/\epsilon_x$, and
$\chi_{S^{y}S^{y}}^i=\sum_{k} (s_{k,i}^{y''}(t)-s_{k,i}^y(t))/\epsilon_y$.

To improve the numerical accuracy of equal-time correlation functions, which are time-independent in the steady state, we additionally average the observables over the time interval $\tau$.

\bigskip

\twocolumngrid

\section{Benchmark of TWA, cumulants, and exact solution}

\begin{figure}
    \centering    \includegraphics[width=\linewidth]{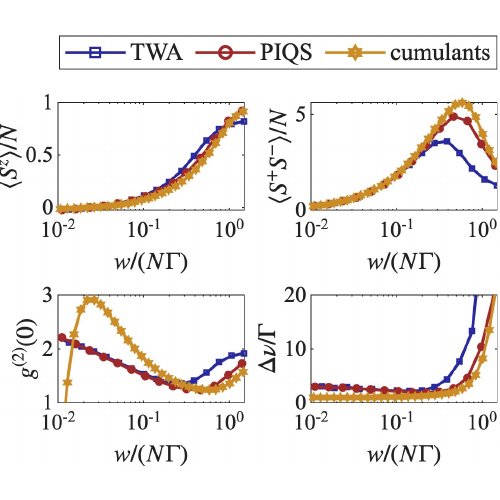}
    \caption{Steady state magnetization, intensity, second order coherence and linewidth calculated for superradiant laser setup containing  $N=40$ spins, evaluated using TWA (blue), exact solution (red) and second order quantum cumulants (yellow).}
    \label{fig:benchmark_40}
\end{figure}

In Fig.~\ref{fig:benchmark_40}, we compare the steady-state solution of the SR laser for $N=40$ spins obtained using the TWA (blue), the exact permutation-invariant quantum solver (PIQS) implemented in the QuTiP library~\cite{PIQS} (red), and a second-order cumulant expansion~\cite{Plankensteiner2022quantumcumulantsjl} (yellow).

The TWA reproduces the exact solution for $w\lesssim \Gamma N/4$, where collective spin loss dominates over local pumping, but develops quantitative errors at stronger pump rates. As the pump increases, the system approaches the fully inverted state, in which the pumping noise vanishes. The semiclassical description does not capture this property exactly because the classical noise does not retain the operator noncommutativity of the underlying quantum dynamics. Consequently, the TWA predicts an incorrect steady state in the regime where the local pump becomes stronger than the collective loss. Nevertheless, the results remain qualitatively correct, although the method overestimates both $g^{(2)}$ and $\Delta\nu$.

The cumulant expansion, by contrast, fails to reproduce some observables at weak pumping and can even yield unphysical negative values of $g^{(2)}$, but performs better in the strong-pump regime. In this regime, the noise associated with local pumping suppresses the correlations generated by collective spin loss, thereby improving the accuracy of the cumulant approximation. The regime in which the cumulant expansion performs best, $w>\Gamma N$, also approaches effectively single-particle behavior as the pump rate increases.

The TWA is therefore particularly suitable for the correlated-pump problem considered here: it is exact in the collective-pump limit and provides a qualitatively reliable description of the steady-state properties in the local-pump limit.

\section{Local spin pump}

As a reference point, we first recall the solution for the standard SR laser with local pump, $\alpha=\infty$. The semiclassical solution for this model via a cumulant expansion was derived in Ref.~\cite{Holland_SR_laser}, which we summarize here.

\begin{figure}
    \centering
    \includegraphics[width=\linewidth]{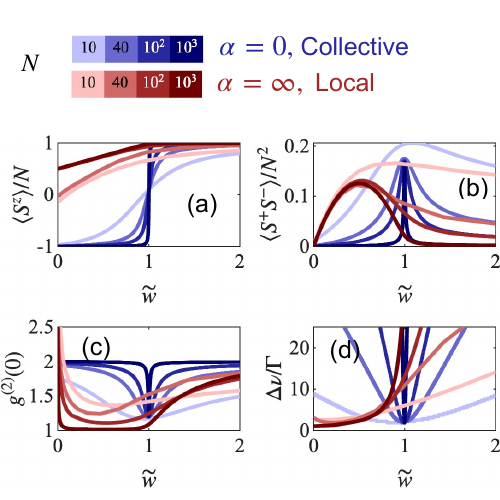}
    \caption{Magnetization (a), intensity (b), second-order coherence (c) and linewidth of the system with collective ($\alpha=0$, blue) and local ($\alpha=\infty$, red) spin pump for $10,~40,~100$ and $10^3$ spins as a function of rescaled pump rate $\tilde w$. Red lines show observables for local spin pump and blue lines correspond to collective spin pump.}
    \label{fig:ed_sol}
\end{figure}

The numerical solution for SR laser is shown in red in Fig.~\ref{fig:ed_sol} as a function of  pump rate $\tilde w=w/(\Gamma N)$ for $\alpha=\infty$. 
Here, SR lasing exists for pumping rates $w\ge\gamma\approx 0$ (the minimal value required for population inversion), up to $w\lesssim{4Ng^{2}}/{\kappa}=N\Gamma$, beyond which the noise accompanying the incoherent pump destroys coherence. In the SR lasing regime, the atoms synchronize and the two-point correlation functions become positive, $\langle \sigma_k^+\sigma_l^-\rangle>0$, which yields $\langle S^+ S^-\rangle=N\left(N-1\right)\langle\sigma_k^{+}\sigma_{l\ne k}^{-}\rangle+{N}\left(1+\langle\sigma^{z}\rangle\right)/2\propto N^2$ and $\langle a^\dagger a\rangle\propto N^2$, while the magnetization can be estimated as $\langle S^z\rangle\simeq w/\Gamma$. The linewidth depends on $w$, reaching its minimal value $\Delta \nu\sim \Gamma+O(1/N)$ at $w\approx \Gamma N/2$, and the second-order coherence is $g^{(2)}\approx 1+O(1/N)$~\cite{Holland_g2}.

The emergence of the intermediate SR lasing phase can be understood from symmetry considerations. At $w=0$, the dynamics is governed by collective dissipation, which obeys strong permutation invariance and weak \textit{collective} $\mathrm{U}(1)$ symmetry. Assuming the spins obey permutation invariance, in the steady state all spins point down and the total magnetization is $\langle S^z\rangle=-N$~\footnote{Note that by itself model~\eqref{eq:model} can host a dark state due to the collective spin loss, but we assume that at time $t=0$ the system was prepared in the spin-polarized state.}. When $w$ becomes nonzero, the steady state is set by the competition between collective spin loss and local spin pump, the latter obeying weak permutation invariance and weak \textit{local} $\mathrm{U}(1)$ symmetry~\cite{buvca2012note}. The weak- and strong-pump regimes are governed by different symmetries, indicating a transition in between. The spin synchronization associated with  breaking of the collective $\mathrm{U}(1)$ symmetry corresponds to the transition from the normal to the lasing regime.

On the mean-field level, the emergence of SR lasing can be viewed as an onset of synchronization~\cite{Holland_Kuramoto,Bruder_laser}. For $\alpha=\infty$, the mean-field dynamics of each spin $\langle \sigma_k^-\rangle$ can be mapped onto the Kuramoto model, which describes the synchronization of the oscillator phases $\phi_k=\mathrm{arg}(\langle \sigma_k^-\rangle)$~\cite{Kuramoto_rev}. When the spins synchronize, $\phi_k=\phi$, the system develops a macroscopic dipole $\langle S^-\rangle \ne 0$, again associated with the spontaneous breaking of the collective $\mathrm{U}(1)$ symmetry. Recalling the connection between the light emitted by the cavity and the macroscopic spin, $a\propto S^-$, one concludes that in the lasing regime $\langle a\rangle\ne 0$, indicating that the light is in a coherent state~\cite{walls2008quantum}. When the pump rate reaches the order $w\propto \Gamma N$ and beyond, it dominates the dynamics and the associated noise destroys synchronization (the system is governed by the local $\mathrm{U}(1)$ symmetry). The steady-state observables are then $\langle S^{x,y}\rangle =0$, $\langle S^z\rangle \to N$, and $g^{(2)}(0)\to 2$, corresponding to thermal light. While this mean-field picture helps build intuition for the origin of the coherent light, we stress that for any exact or even semiclassical method working at a finite $N$, symmetry considerations imply $\langle S^-\rangle =0$~\cite{kirton2018superradiant}, so that all information is instead carried by the two-point and higher-order correlation functions.

\section{Collective spin pump}
 In the $\alpha=0$ case all atoms are pumped collectively.  The dynamics are exactly solvable within the symmetric Dicke manifold with fixed total spin $S=N/2$, and have been partially reported, for instance, in Refs.~\cite{reilly2025fullycollectivesuperradiantlasing,shimshi2024dissipativephasetransitionmetrology,chelpanova2026knobtuneallphasecontrolled}. When both pump and loss are collective, they preserve permutation symmetry, so that an initially symmetric state remains confined to the Dicke manifold with fixed total spin $S$, $\rho(t)=\sum_m \alpha_m(t)\,|m\rangle\langle m|$. The steady state follows from detailed balance, $\alpha_{m+1}=({w}/{\Gamma})\,\alpha_m=r\alpha_m=r^{m+1}\alpha_{-S}$ with $\sum_{m=-S}^{S}\alpha_m=1$, allowing all observables to be obtained analytically (note that for the local spin pump, detailed balance breaks down inside the SR lasing phase, cf. Ref.~\cite{shankar2026dynamicalaspectssteadystatesubradiance}). The steady state can be described by the Gibbs ensemble $\rho=\exp(-\beta H)/Z$, with $H=\omega S^z/2$ and effective temperature $\beta\omega=\ln(\Gamma/w)$. For $\omega>0$ it therefore corresponds to a positive-temperature state for $\Gamma>w$, a negative-temperature, population-inverted state for $w>\Gamma$, and an infinite-temperature state  (within a given Dicke manifold) at $w=\Gamma$.

For finite $w\ne\Gamma$, the emitted light is thermal, with $g^{(2)}\to 2$ and linewidth $\Delta\nu=N|w-\Gamma|$. At $w=\Gamma$, the minimal value $g^{(2)}=6/5$ is reached, while $\langle S^+ S^-\rangle \propto N^2$ and the linewidth remains finite, $\Delta\nu=\Gamma+w=2\Gamma$. We call this regime \textit{partially coherent SR lasing}. From the symmetry perspective, for $\alpha=0$ the master equation (1) preserves the  weak collective $\mathrm{U}(1)$ symmetry, preventing the development of a macroscopic dipole $\langle S^-\rangle \ne 0$ (note that the mean-field solution for the collective model also yields $\langle S^{-}\rangle=0$). Consequently, the system does not undergo a conventional lasing transition and remains described by a thermal state, with $\langle a\rangle =0$ and $g^{(2)}=2$. An exception occurs at $w=\Gamma$, where the Liouvillian becomes additionally invariant under $S^+\leftrightarrow S^-$. Remarkably, this coincides with the emergence of the ultra-narrow linewidth $\Delta\nu=2\Gamma$, even though the emitted light is only partially coherent. It is also worth noting that the collective spin pump preserves the total spin and cannot counteract local spin loss, so that in reality, due to the inescapable presence of local spontaneous emissions, such a system would relax to a trivial steady state with all spins pointing down. Thus, for the two-level system, it is crucial that pump and collective loss respect different symmetries in order to leave room for spontaneous symmetry breaking and nontrivial steady states. We plot exact solution in Fig.~\ref{fig:ed_sol} (blue lines) for different system sizes as a function of  pump rates $\tilde w=w/\Gamma.$ Here, as system size increases, the ultra-narrow regime collapses to a single point at  $w=\Gamma$.

On the other hand, the naive advantage of the collective spin pump is that at the optimal point $w=\Gamma$ the pump rate does not scale with system size (compare with $w=\Gamma N/2$ for optimal SR lasing with local pump), which means that the intensity of auxiliary lasers mediating such pumping need not be extensive. As we show in the main text,  for finite $\alpha$ the correlated pump breaks strong permutation invariance and the system explores a larger Hilbert space, allowing the generation of coherent ultra-narrow light at intermediate values of $\alpha$ for pump rates that scale slower than linearly with $N$.

\begin{figure}
    \centering
    \includegraphics[width=\linewidth]{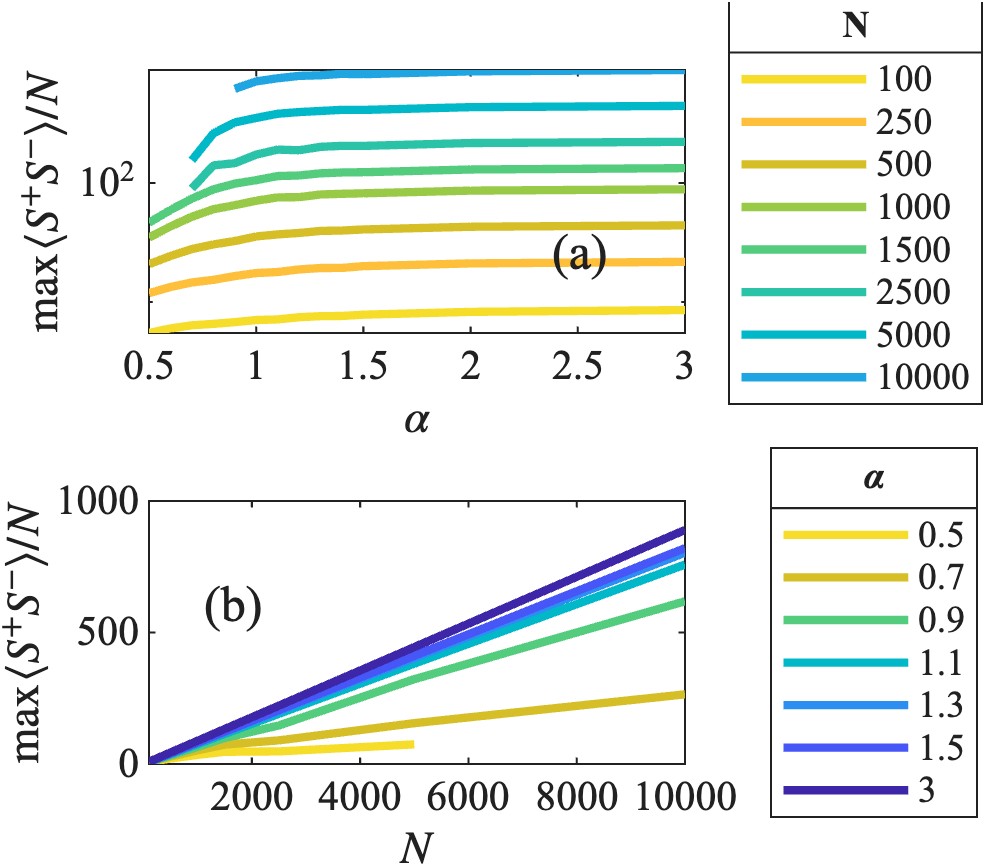}
    \caption{Scaling of the normalized maximal intensity $\max \langle S^+ S^-\rangle/N $ as a function of $\alpha$ (a) for different $N$ (indicated by color) and (b) $N$ for different values of exponents $\alpha$.}
    \label{fig:intensity}
\end{figure}

\section{Scaling of the intensity}

In this section, we provide additional data on the scaling of the maximal intensity with the exponent $\alpha$ and the system size; see Fig.~\ref{fig:intensity}. We follow the same protocol as in Fig.~3 of the main text, preparing the system in the steady state for $10$ equally spaced values of $\tilde w$ between $\tilde w=1/10$ and $\tilde w=1$. For the larger system sizes, $N\ge 1000$, we evaluate only a few points in the vicinity of the optimal lasing region.

Panel (a) shows the dependence of the normalized maximal intensity on $\alpha$ and indicates saturation in the range $1<\alpha<2$. Panel (b) shows the dependence of the maximal intensity on the system size. For all values of $\alpha$, the data suggest
$\max\langle S^+S^-\rangle/N\propto N$, corresponding to the superradiant scaling $\max\langle S^+S^-\rangle\propto N^2$. Such scaling is characteristic of emission mediated by collective spin loss. For fixed $N$, the intensity increases with $\alpha$ and approaches the value obtained in the local-pump limit.

We consider only finite values $\alpha>0.5$, since the relaxation time diverges as $1/(w\alpha)$ near the collective limit, making simulations at smaller $\alpha$ prohibitively time-consuming. We also exclude $\alpha=0$ from this figure. Although this limit is readily accessible numerically, strong permutation symmetry confines the dynamics to the symmetric $\mathrm{SU}(2)$ sector, making the corresponding observables qualitatively different from those obtained in the limit $\alpha\to 0^+$.

\end{document}